\documentclass[twocolumn,showpacs,preprintnumbers,amsmath,prl,amssymb,superscriptaddress]{revtex4}

\pdfoutput=1

\usepackage{graphicx}
\usepackage{dcolumn}
\usepackage{bm}

\def\mwimp{\rm{m_{\chi}}}
\def\csnospin{\rm{\sigma_{\chi N}^{SI}}}
\def\effbs{\epsilon _{\rm BS}}
\def\lmbdbs{\lambda _{\rm BS}}
\def\pge{{\it p}{\rm Ge}}
\def\nge{{\it n}{\rm Ge}}

\begin{document}


\title{
Limits on spin-independent 
couplings of WIMP dark matter
with a p-type point-contact
germanium detector
}

%

\newcommand{\as}{Institute of Physics, Academia Sinica, 
Taipei 11529, Taiwan.}
\newcommand{\thu}{Department of Engineering Physics, Tsinghua University,
Beijing 100084, China.}
\newcommand{\metu}{Department of Physics,
Middle East Technical University, Ankara 06531, Turkey.}
\newcommand{\deu}{Department of Physics,
Dokuz Eyl\"{u}l University, Buca, \.{I}zmir 35160, Turkey.} 
\newcommand{\ciae}{Department of Nuclear Physics,
Institute of Atomic Energy, Beijing 102413, China.}
\newcommand{\bhu}{Department of Physics, Banaras Hindu University,
Varanasi 221005, India.}
\newcommand{\nku}{Department of Physics, Nankai University,
Tianjin 300071, China.}
\newcommand{\scu}{Department of Physics, Sichuan University,
Chengdu 610065, China.}
\newcommand{\ks}{Kuo-Sheng Nuclear Power Station,
Taiwan Power Company, Kuo-Sheng 207, Taiwan.}
\newcommand{\corr}{htwong@phys.sinica.edu.tw} 

\author{ H.B.~Li }  \affiliation{ \as }
\author{ H.Y.~Liao }  \affiliation{ \as }
\author{ S.T.~Lin }  \affiliation{ \as } \affiliation{ \deu }
\author{ S.K.~Liu } \affiliation{ \scu }
\author{ L.~Singh }  \affiliation{ \as } \affiliation{ \bhu }
\author{ M.K.~Singh }  \affiliation{ \as } \affiliation{ \bhu }
\author{ A.K.~Soma }  \affiliation{ \as } \affiliation{ \bhu }
\author{ H.T.~Wong } \altaffiliation[Corresponding Author: ]{ \corr } \affiliation{ \as }
\author{ Y.C.~Wu } \affiliation{ \thu }
\author{ W.~Zhao }  \affiliation{ \thu } 
\author{ G.~Asryan }  \affiliation{ \as }
\author{ Y.C.~Chuang }  \affiliation{ \as }
\author{ M.~Deniz } \affiliation{ \deu }
\author{ J.M.~Fang } \affiliation{ \ks }
\author{ C.L.~Hsu }  \affiliation{ \as }
\author{ T.R.~Huang }  \affiliation{ \as }
\author{ G.~Kiran Kumar } \affiliation{ \as }
\author{ S.C.~Lee }  \affiliation{ \as }
\author{ J.~Li }  \affiliation{ \thu } 
\author{ J.M.~Li }  \affiliation{ \thu } 
\author{ Y.J.~Li }  \affiliation{ \thu } 
\author{ Y.L.~Li }  \affiliation{ \thu } 
\author{ C.W.~Lin }  \affiliation{ \as }
\author{ F.K.~Lin }  \affiliation{ \as }
\author{ Y.F.~Liu } \affiliation{ \as } \affiliation{ \nku }
\author{ H.~Ma } \affiliation{ \thu }
\author{ X.C.~Ruan } \affiliation{ \ciae }
\author{ Y.T.~Shen }  \affiliation{ \as }
\author{ V.~Singh }  \affiliation{ \bhu }
\author{ C.J.~Tang } \affiliation{ \scu }
\author{ C.H.~Tseng }  \affiliation{ \as }
\author{ Y.~Xu } \affiliation{ \as } \affiliation{ \nku }
\author{ S.W.~Yang }  \affiliation{ \as }
\author{ C.X.~Yu } \affiliation{ \as } \affiliation{ \nku }
\author{ Q.~Yue } \affiliation{ \thu }
\author{ Z.~Zeng }  \affiliation{ \thu } 
\author{ M.~Zeyrek } \affiliation{ \metu }
\author{ Z.Y.~Zhou } \affiliation{ \ciae }

\collaboration{TEXONO Collaboration}



\date{\today}

\begin{abstract}

We report new limits on spin-independent 
WIMP-nucleon interaction cross-section
using 39.5~kg-days of data taken with 
a p-type point-contact germanium detector
of 840~g fiducial mass 
at the Kuo-Sheng Reactor Neutrino Laboratory.
Crucial to this study is 
the understanding of the selection procedures
and, in particular, the bulk-surface events differentiation
at the sub-keV range.
The signal-retaining and background-rejecting
efficiencies were measured with 
calibration gamma sources and a novel n-type
point-contact germanium detector.
Part of the parameter space in 
cross-section versus WIMP-mass
implied by various experiments 
is probed and excluded.

\end{abstract}

\pacs{
95.35.+d,
29.40.-n,
98.70.Vc
}
\keywords{
Dark matter, 
Radiation Detector,
Background radiation 
}

\maketitle

About one quarter of
the energy density of the universe 
can be attributed to
Cold Dark Matter\cite{cdmpdg12}, 
whose nature and properties are unknown.
Weakly Interacting Massive Particles 
(WIMP, denoted by $\chi$) are
its leading candidates.
There are intense experimental efforts
to study $\chi$N$\rightarrow$$\chi$N elastic scattering
via the direct detection of nuclear recoils.
Most experimental programs are optimized 
for mass range at $\mwimp$$\sim$10$-$100~GeV,
motivated by popular supersymmetric models.
Germanium detectors sensitive to sub-keV recoil energy
were identified and demonstrated as possible means to
probe the ``low-mass'' WIMPs
with $\mwimp$$<$10~GeV\cite{ulege}.
This inspired development of p-type point-contact
germanium detectors ($\pge$) 
with modular mass of kg-scale\cite{ppcge}.

Our earlier measurements 
at the Kuo-Sheng Reactor Neutrino Laboratory 
(KSNL, with a shallow depth of about 
30~meter-water-equivalence) 
using a 4-element array with
a total mass of 20~g and analysis 
threshold of 220~eVee 
(``ee'' denoting electron-equivalence energy throughout) 
have placed constraints on 
$\mwimp$$>$3~GeV\cite{texonocdm07}.
The CoGeNT experiment reported data 
with a 440~g detector\cite{cogent}, 
showing an excess of events 
at the sub-keV range over
the background models. 
A consistent annual modulation signature was observed.
Allowed region in the spin-independent
$\chi N$ couplings ($\csnospin$) was
derived. Intense interest and 
theoretical speculations 
in the low-mass WIMP region 
were generated\cite{lowimptheory}.
The low energy data of 
the CDMS and XENON experiments\cite{cdmsxenonlow}
have subsequently excluded the allowed region
with different detector techniques, 
while the original interpretations 
were defended\cite{cogentreply}. 
It is crucial to have independent experiments
which can probe the CoGeNT allowed region and 
provide further understanding on 
the detector response and the nature of 
the sub-keV events in Ge-detectors.

We report new results 
with a $\pge$ of 840~g fiducial mass 
(actual crystal mass 926~g) at KSNL.
The low-background facilities as well as the hardware, 
trigger and data acquisition configurations
were described in our previous work\cite{texonocdm07,ksnlneutrino}. 
The detector was enclosed by an NaI(Tl)
anti-Compton (AC) detector and copper passive shieldings
inside a plastic bag purged by nitrogen gas evaporated from
the liquid nitrogen dewar. This set-up was further 
shielded by, from inside out, 5~cm of copper,
25~cm of boron-loaded polyethylene, 5~cm of steel
and 15~cm of lead. This structure was surrounded by 
cosmic-ray (CR) veto panels made of 
plastic scintillators read out by photomultipliers. 
Both AC and CR detectors are crucial,
serving both as vetos to reject background 
and as tags to identify samples for
efficiency measurements.

Signals from the point-contact 
are supplied through a reset preamplifier.
The output is distributed to 
a fast-timing amplifier which 
keeps the rise-time information, and to
amplifiers at both 6~$\mu$s 
and 12~$\mu$s shaping time 
which provide energy information. 
Signals from the outer surface-electrode 
are processed with a resistive feedback preamplifier
and followed by amplifier at 4~$\mu$s shaping time.
The fast-timing, slow-shaping and AC-NaI(Tl) output 
were digitized by flash analog-to-digital converters
at 200~MHz, 60~MHz and 20~MHz, respectively.
The discriminator and timing outputs of the CR panels
were also recorded.
The physics triggers are provided by the 
discriminator output of the 6~$\mu$s shaping pulses.  
The trigger efficiency of 100\% above 300~eVee
was verified by test pulser events.
A total of 53.8~days of data were taken,
where the data acquisition dead time 
was 12.6\%, measured
by random trigger events.
Energy calibration was achieved by the internal
X-ray peaks and the zero-energy was defined with 
the pedestals provided by the random events.
The range in between was cross-checked 
with pulser events.
The electronics noise-edge is at 400~eVee.


\begin{figure}[t] 
\includegraphics[width=8.5cm,height=5cm]{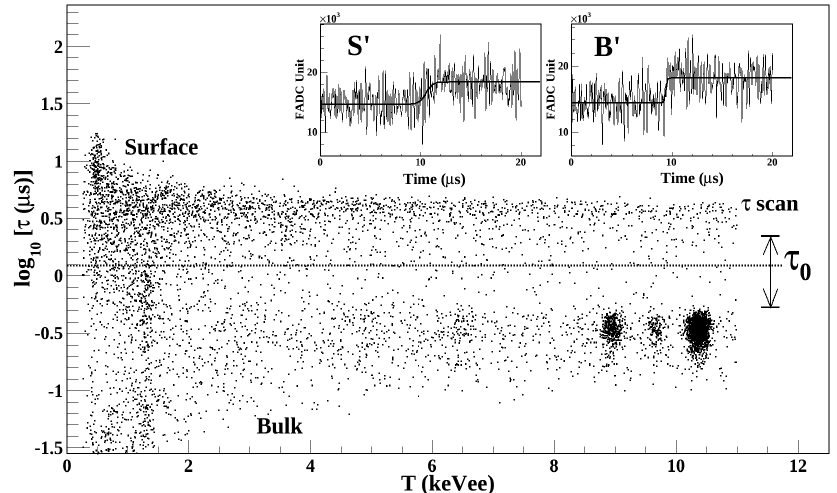}\\
\caption{
Scatter plot of the $\pge$ rise time (${\rm log_{10}}[ \tau]$) 
versus energy. The $\tau_0$-line corresponds 
to the BS cut in this analysis, with 
$\tau$-scan indicating the range of cut-stability test. 
Typical B$'$(S$'$) pulses at {\it T}$\sim$700~eVee are depicted
in the insets.
}
\label{fig::bscut}
\end{figure}

A cut-based analysis was adopted.
There are three categories of selection criteria:
(i) the ``physics versus noise events'' 
(PN) cuts differentiate physics signals
from spurious electronic noise;
(ii) the AC and CR cuts identify events 
with activities only at the $\pge$ target; and 
(iii) the ``bulk versus surface events''
(BS) cut selects events at the interior.
In addition, the efficiencies and 
suppression factors ($\epsilon_{X}$,$\lambda_{X}$)
for every selection ($X$=PN,AC,CR,BS) are measured.
They correspond to the probabilities of (signal,background) 
events being correctly identified.
The physics events selected by the PN cuts
are categorized by 
``AC$^{-(+)}$$\otimes$CR$^{-(+)}$$\otimes$B(S)'', where
AC$^{-(+)}$ and CR$^{-(+)}$ represent
AC and CR signals in 
anti-coincidence(coincidence), respectively,
while B(S) denote the bulk(surface) samples. 
The $\chi$N candidates would therefore
manifest as AC$^-$$\otimes$CR$^-$$\otimes$B events.

\begin{figure}[t] 
{\bf  (a)}
\includegraphics[width=8.5cm,height=4cm]{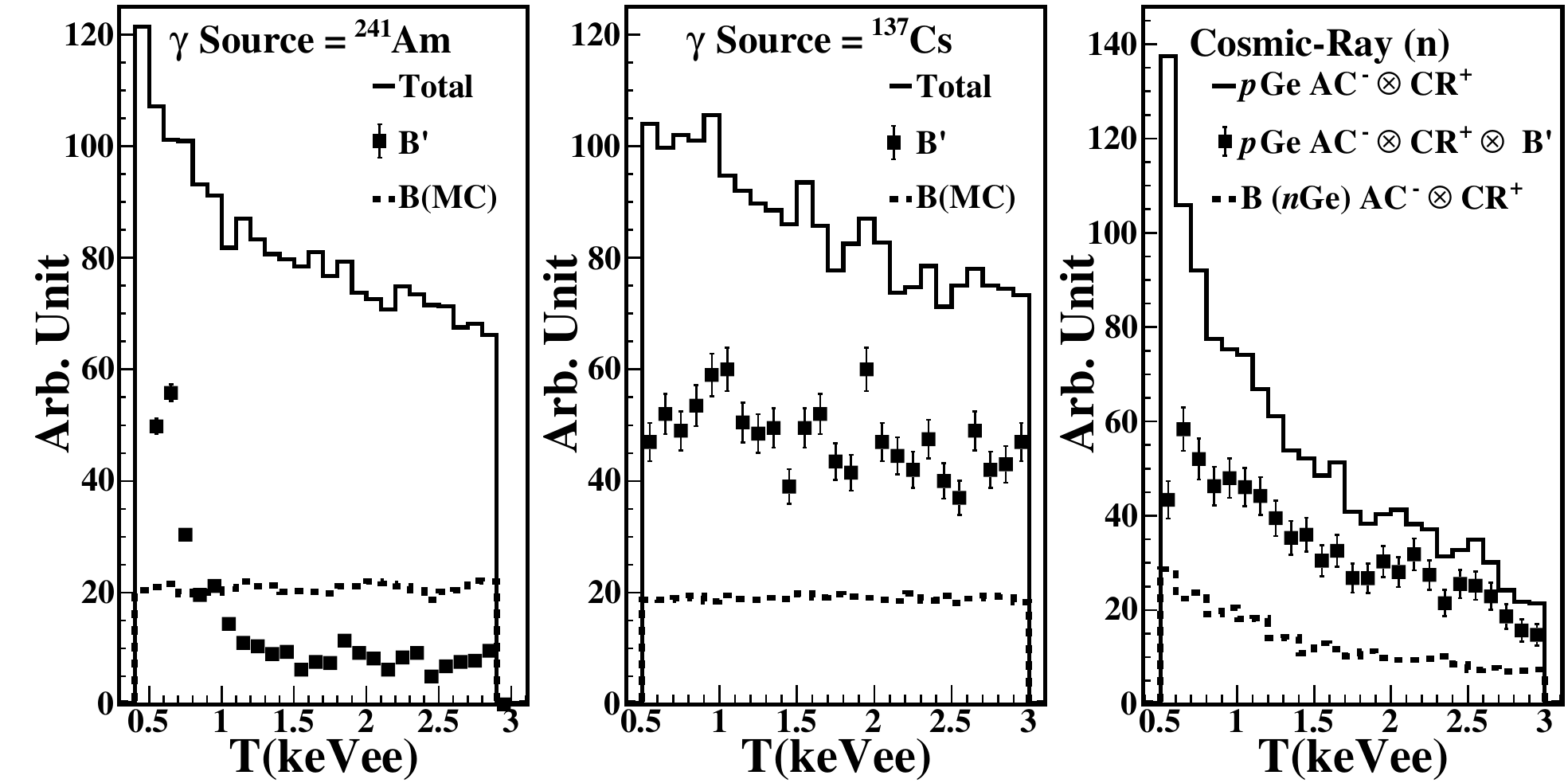}
{\bf (b)}\\
\includegraphics[width=8.5cm,height=4cm]{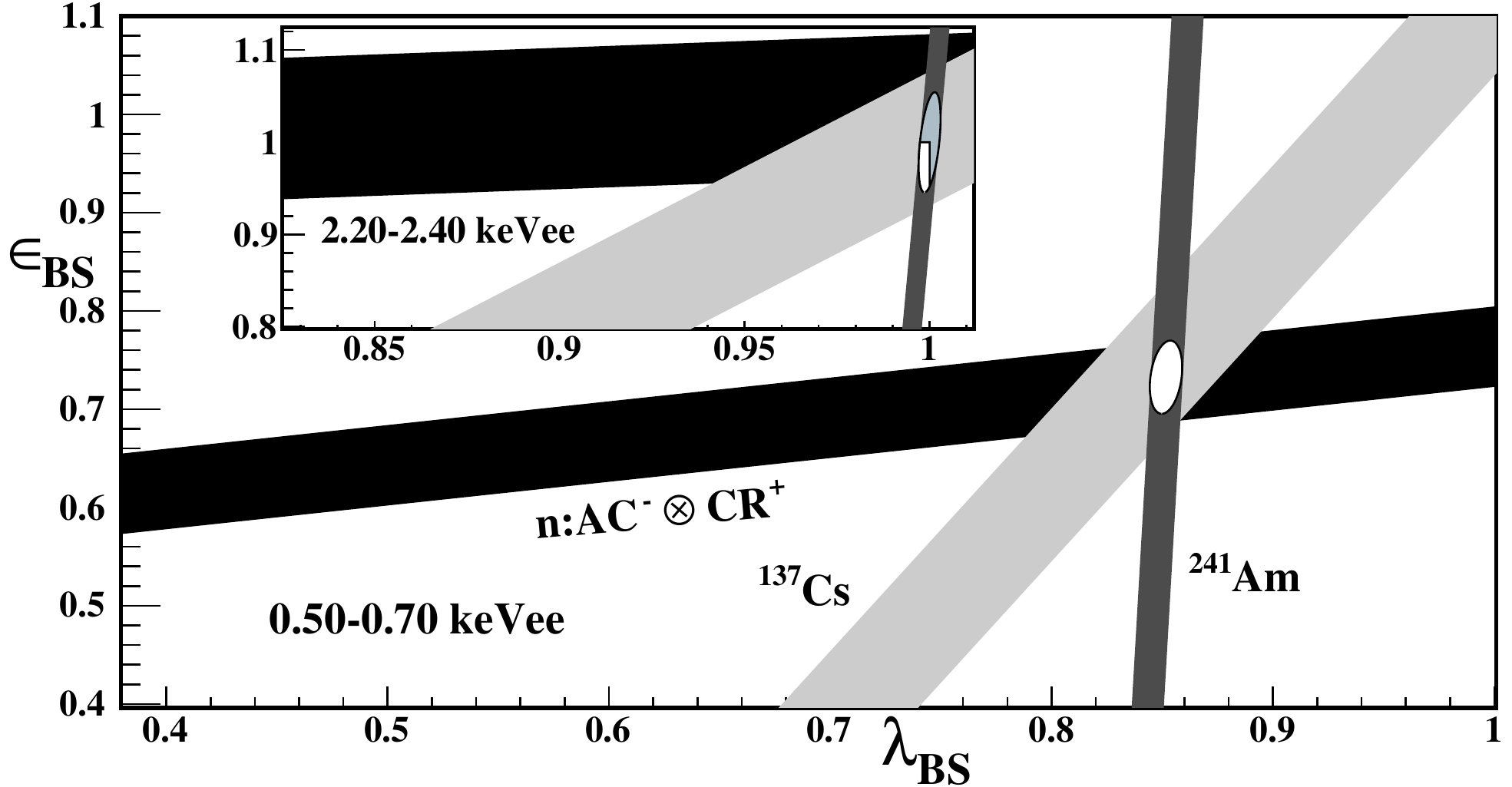}
{\bf (c)}\\
\includegraphics[width=8.5cm,height=4cm]{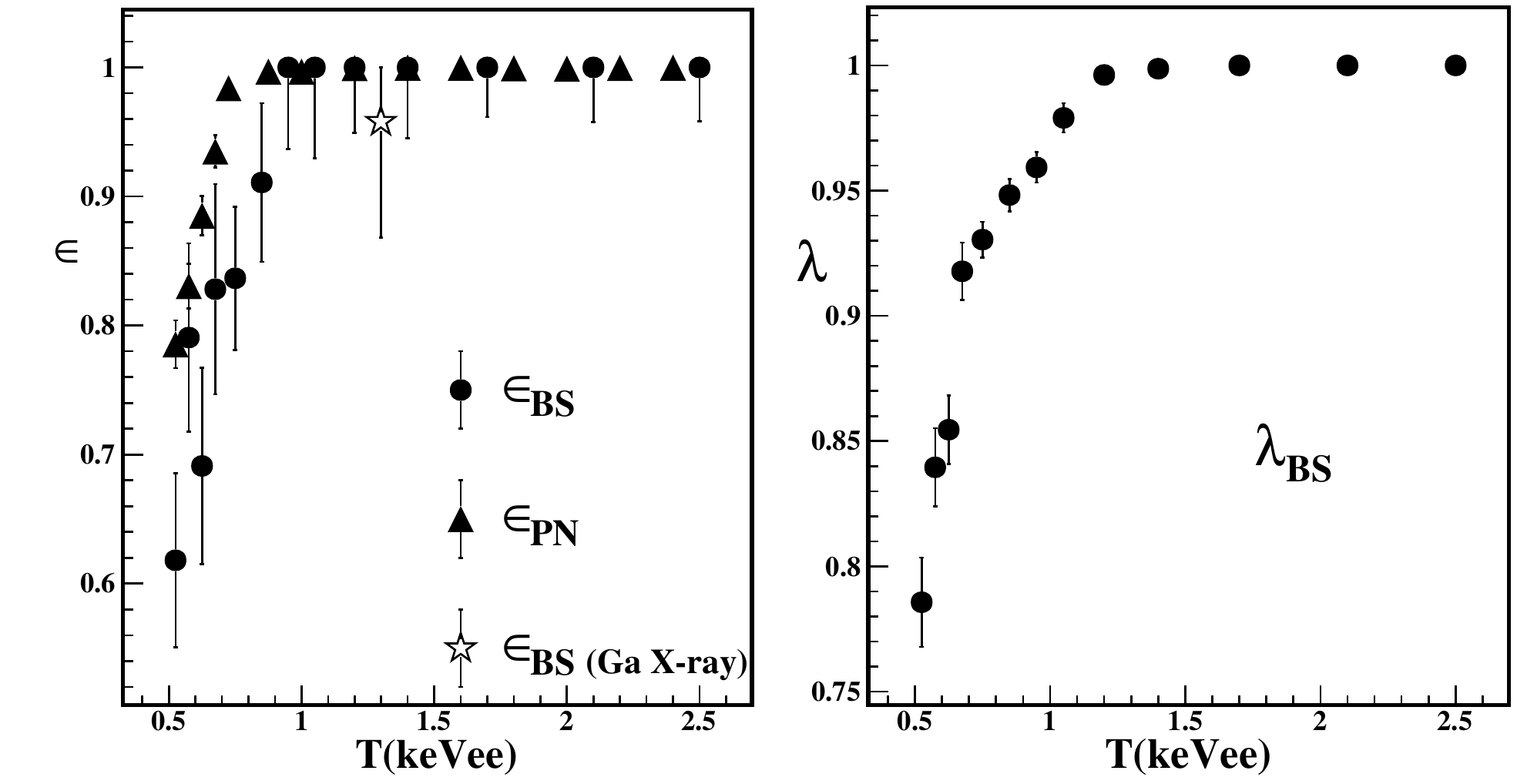}
\caption{
The derivation of 
$( \effbs , \lmbdbs )$$-$
(a) The measured Total and B$'$ spectra from $\pge$
with the surface-rich 
$\gamma$-ray ($^{241}$Am,$^{137}$Cs)  
and bulk-rich cosmic-ray induced neutrons. 
They are compared to reference B-spectra 
acquired via simulations for $\gamma$-rays
and $\nge$ measurement for cosmic-neutrons.
(b) Allowed bands at threshold and at a high energy band.
(c) The measured ($\effbs$,$\lmbdbs$)
and $\epsilon_{PN}$ as functions of energy. 
Independent measurement on $\effbs$ with
Ga-L X-rays is included. 
}
\label{fig::bscalib}
\end{figure}

Background suppression with the PN, AC and CR cuts 
and the evaluations of their respective 
$( \epsilon_X , \lambda_X )$
follow the well-studied procedures of
earlier experiments\cite{ksnlneutrino,texonocdm07,geqf}.
The PN cuts are based on pulse shape characteristics and
correlations among the fast and shaping signals.
They suppress spurious triggers induced by 
microphonics effects or 
the tails of pedestal fluctuations. 
Background induced by the preamplifier reset is
identified by the timing correlations
with the reset instant.
The {\it in situ} doubly-tagged AC$^+$$\otimes$CR$^+$ events
serve as the physics reference samples,
with which 
$\epsilon_{PN}$ shown 
in Figure~\ref{fig::bscalib}c
are accurately measured. 
The majority of the electronics-induced events
above noise-edge are identified 
($\lambda_{PN}$$\sim$1).
The efficiencies for AC and CR selections
are measured by the random events to be, 
respectively,
$\epsilon _{\rm AC}$$>$0.99 and
$\epsilon _{\rm CR}$=0.93.
The suppressions are $\lambda _{\rm AC}$=1.0
above the NaI(Tl) threshold of 20~keVee, while
$\lambda _{\rm CR} = 0.92$, measured by
reference cosmic samples in which 
the energy depositions at 
NaI(Tl) are above 20~MeVee.

The BS selection, on the other hand,
is a unique feature to $\pge$.
The surface-electrode is a
lithium-diffused n$^+$ layer of mm-scale thickness.
Partial charge collection
in the surface layer gives rise to 
reduced measureable energy and slower rise-time ($\tau$)
in its fast-timing output, 
as compared to those in the
bulk region\cite{gesurface,cogent,pcgerandd}.
The thickness of the S layer
was derived to be (1.16$\pm$0.09)~mm, 
via the comparison of
simulated and observed intensity ratios
of $\gamma$-peaks from 
a $^{133}$Ba source\cite{ppcmj12}.
This gives rise to a fiducial mass
of 840~g, or a data size of 39.5~kg-days.

The $\rm{log_{10}} [ \tau ]$ versus measured energy({\it T}) 
scatter plot is displayed in Figure~\ref{fig::bscut}. 
The boundary between the bulk and surface layers 
is not well defined, giving rise to events between
the two bands.
The observed and actual rates are denoted by
(B$'$,S$'$) and (B,S), respectively.
Events with $\tau$ less(larger) than $\tau_0$ are 
categorized as B$'$(S$'$). 
Typical B$'$(S$'$) events at {\it T}$\sim$700~eVee are shown. 
At{\it T}$>$2.7~keVee
where the $\tau$-resolution is better than the
separation between the two bands,
the assignments B=B$'$ and S=S$'$ are justified. 
At lower energy, (B$'$,S$'$) and (B,S)
are related by the coupled equations:
\begin{eqnarray}
B' & = &  \effbs \cdot B ~  +  ~ ( 1 - \lmbdbs ) \cdot S \\
S' & = &  ( 1 - \effbs) \cdot B ~  +  ~ \lmbdbs \cdot S ~~ , \nonumber
\label{eq::elcoupled}
\end{eqnarray}
with an additional unitarity constrain:
B+S=B$'$+S$'$.

\begin{figure}[t] 
\includegraphics[width=8.5cm,height=6.5cm]{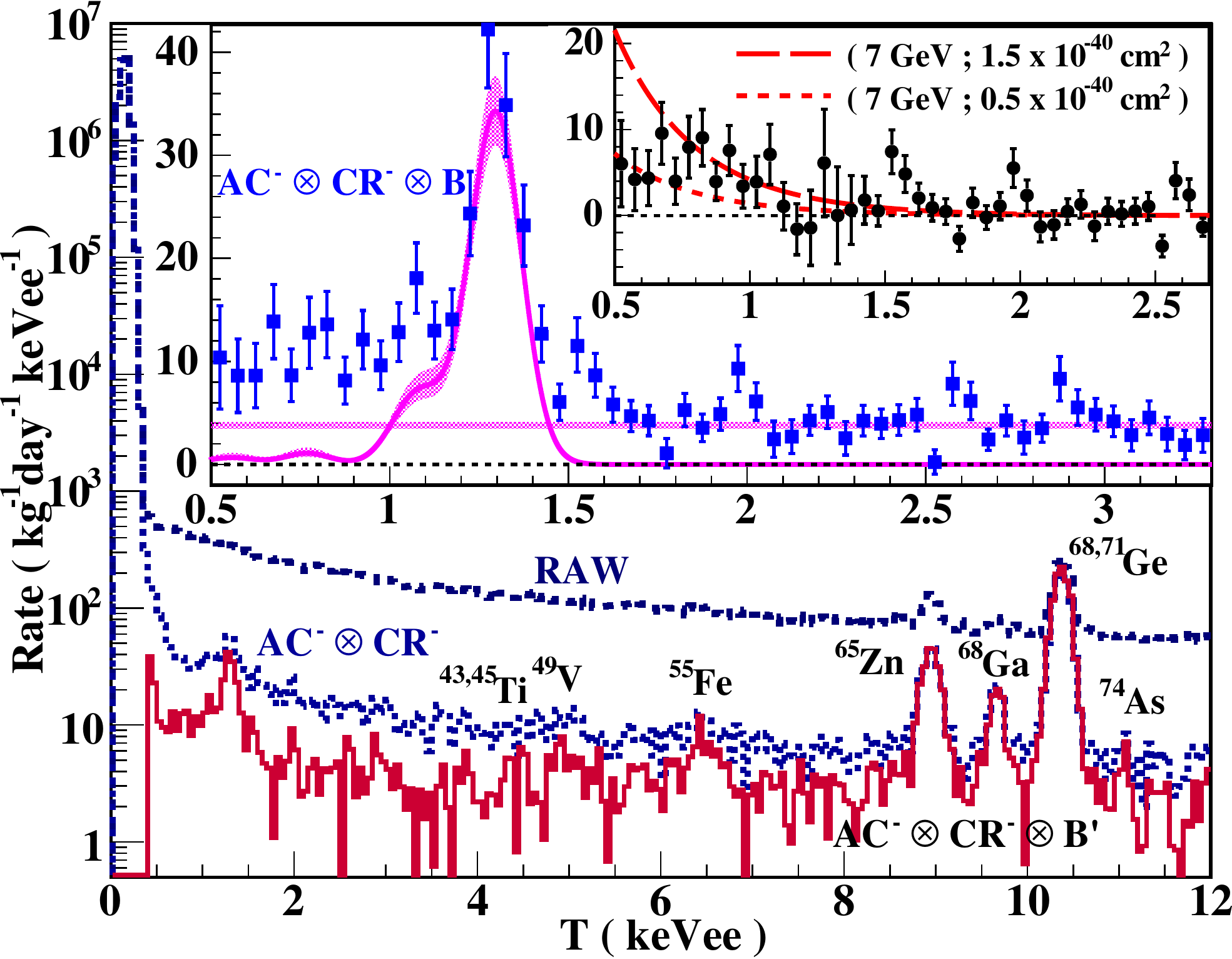}
\caption{
Measured energy spectra, showing the
raw data and those with AC$^-$$\otimes$CR$^-$($\otimes$B$'$) selections. 
The large inset shows the 
($\effbs$,$\lmbdbs$)-corrected AC$^-$$\otimes$CR$^-$$\otimes$B spectrum, 
with a flat background and L-shell X-ray peaks overlaid.
The small inset depicts
the residual spectrum superimposed with 
that due to an allowed (excluded)
cross-section at $\mwimp$=7~GeV.
}
\label{fig::spectra}
\end{figure}

The calibration of ($\effbs$,$\lmbdbs$) 
involves at least two measurements of (B$'$,S$'$)
where (B,S) are independently known. 
The pulser events are inappropriate 
since their fast-timing output 
exhibit different pulse shapes from those
of physics events.
Instead three complementary data samples,
as displayed in Figure~\ref{fig::bscalib}a, 
were adopted:

{\bf (I)} 
Surface-rich events with $\gamma$-ray sources $-$
Calibrations with both 
low and high energy $\gamma$-sources 
($^{241}$Am at 60~keVee and $^{137}$Cs at 662~keVee, respectively)
were performed. 
As displayed in Figure~\ref{fig::bscalib}a,
the measured B$'$-spectra
are compared to the reference B
derived from full simulation 
with surface layer thickness of 1.16~mm as input.
The simulated B-spectra due to external $\gamma$-sources 
over a large range of energy
are flat for {\it T}$<$10~keVee. 

{\bf (II)} 
Bulk-rich events with cosmic-ray induced fast neutrons $-$
A 523~g first-of-its-kind
n-type point-contact 
germanium ($\nge$) detector was constructed. 
The components and dimensions are identical 
to those of $\pge$. 
The surface of $\nge$
is a p$^+$ boron implanted electrode of 
sub-micron thickness.
There are no anomalous surface effects.
Data were taken under identical shielding configurations
at KSNL.
The trigger efficiency was 100\% above T=500~eVee, 
and energy calibration was obtained from the
standard internal X-ray lines.
The AC$^-$$\otimes$CR$^+$ condition selects 
cosmic-ray induced fast neutron events 
without associated $\gamma$-activities,
which manifest mostly($\sim$85\%) as bulk events.
Accordingly, the AC$^-$$\otimes$CR$^+$ spectrum in $\nge$
is taken as the B-reference
and compared with those of 
AC$^-$$\otimes$CR$^+$$\otimes$B$'$ in $\pge$.

Using calibration data (I) and (II),
($\effbs$,$\lmbdbs$) are measured by 
solving the coupled 
equations in Eq.~\ref{eq::elcoupled}.
Standard error propagation formulae are adopted 
to derive their uncertainties 
using errors in (B,B$'$,S$'$) as input.
As examples,
the three allowed bands  
at threshold and at a high energy band
are illustrated in Figure~\ref{fig::bscalib}b.
The different orientations of the bands are 
consequences of the different depth distributions 
of the samples, which give rise to different 
B:S ratios.
The bands have common overlap regions, indicating
the results are insensitive to the event locations.
The surface-rich $\gamma$-events 
and the bulk-rich cosmic-ray induced neutron-events
play complementary roles in constraining
$\lmbdbs$ and $\effbs$, respectively.
The results are depicted in Figure~\ref{fig::bscalib}c,
with $\epsilon_{PN}$ overlaid.
By comparing the measured 
{\it in situ} Ga-L X-ray peak at 1.3~keVee 
after BS-selection to that predicted by the corresponding
K-peak at 10.37~keVee, 
a consistent $\effbs$ is independently measured.

The raw spectrum and those 
of AC$^-$$\otimes$CR$^-$($\otimes$B$'$) 
are depicted in Figure~\ref{fig::spectra}.
The peaks correspond to known K-shell X-rays 
from the cosmogenically-activated isotopes.
The ($\effbs$,$\lmbdbs$)-corrected spectrum 
of AC$^-$$\otimes$CR$^-$$\otimes$B is shown in the large inset. 
Errors above {\it T}$\sim$800~eVee 
are dominated by statistical uncertainties, 
while those below have additional contributions
from the BS calibration errors of Figure~\ref{fig::bscalib}c,
which increase as the efficiencies deviate
from unity at low energy.
The analysis threshold is placed at 500~eVee,
where ($\effbs$,$\lmbdbs$)$\sim$0.5
and the BS selection is no longer valid.
The stability of 
($\epsilon_{BS}$,$\lambda_{BS}$,B$'$,S$'$,B) 
is studied over changes of $\tau_0$
within the $\tau$-scan range of Figure~\ref{fig::bscut}. 
Measurements of B are stable and independent of $\tau_0$,
as indicated by the small variations 
relative to the uncertainties.
On the contrary, ($\epsilon_{BS}$,$\lambda_{BS}$) 
exhibit significant shifts in the expected directions.
These features indicate that the BS calibration procedures
are valid and robust. The systematic errors 
due to parameter choices are of minor effects 
to the total uncertainties.

\begin{figure}[t] 
\includegraphics[height=7.2cm,width=8.5cm]{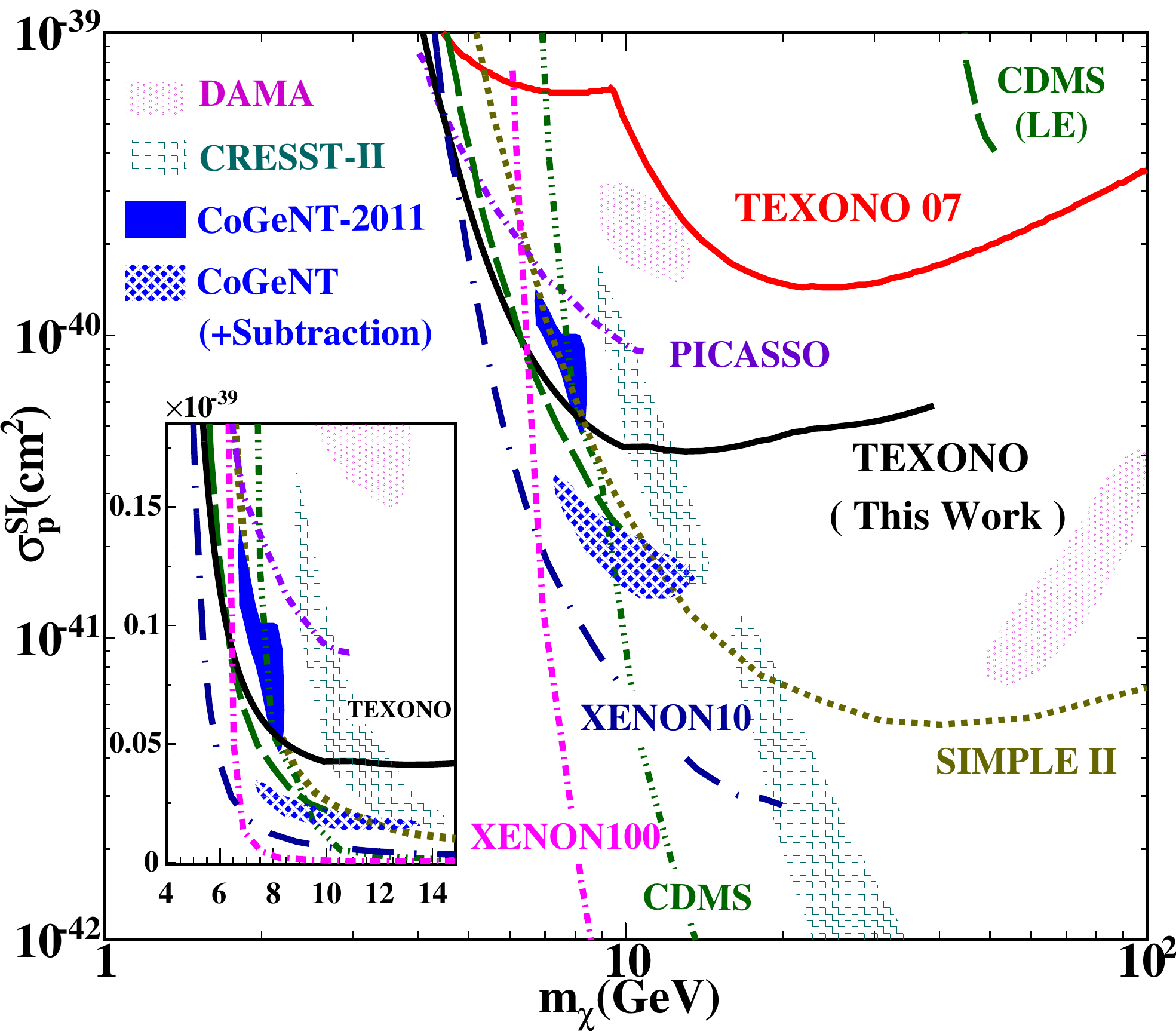}
\caption{
Exclusion plot of spin-independent $\chi$N coupling
at 90\% confidence level,
superimposed with the results 
from other benchmark experiments
and CoGeNT with and without
surface background subtraction.
}
\label{fig::explot}
\end{figure}

High energy $\gamma$-rays from ambient radioactivity 
produce flat electron-recoil background at low energy,
as verified by the $^{241}$Am and
$^{137}$Cs spectra of Figure~\ref{fig::bscalib}a, 
and by the {\it in situ} AC$^+$$\otimes$CR$^-$$\otimes$B spectra. 
This, together with the L-shell X-ray lines predicted
by the higher energy K-peaks, 
are subtracted from AC$^-$$\otimes$CR$^-$$\otimes$B. 
At a given $\mwimp$, the flat background 
is measured at an energy range of at least 1.7~keVee and 
beyond the tail ($<$1\%) of the $\chi$N recoil spectra.
The residual spectrum corresponds to $\chi$N candidate events
and is depicted in the small inset of Figure~\ref{fig::spectra}. 
Constraints on $\csnospin$
are derived via the ``binned Poisson'' method\cite{bpoisson}
with conventional astrophysical models\cite{cdmpdg12}
(local density of 0.3~GeV/cc and 
Maxwellian velocity distribution with
$v_0$=220~km/s and $v_{esc}$=544~km/s).
The event rates of $\chi$N spin-independent
interaction cannot be larger 
than the residual spectrum.
The quenching function in Ge is
derived with the TRIM software 
which matches well with existing data\cite{geqf}.
As illustration,
$\chi$N recoil spectrum due to
an allowed (excluded) 
$\csnospin$ at $\mwimp$=7~GeV 
is shown in Figure~\ref{fig::spectra}.
Exclusion plot of $\csnospin$ 
versus $\mwimp$ at 90\% confidence level
is displayed in Figure~\ref{fig::explot}.
Bounds from other benchmark 
experiments are superimposed\cite{cogent,cdmsxenonlow,keycdmexpt}.
The favored region from the CoGeNT data
with additional surface background 
subtraction\cite{lowimptheory} is included.
An order of magnitude improvement 
over our previous results\cite{texonocdm07}
is achieved. 
Part of the published DAMA, CRESST-II and
CoGeNT allowed regions
are probed and excluded.
We note that an excess remains
in the sub-keV region 
not yet accounted for in this analysis,
the understanding of which is the theme
of our on-going investigations.

Studies continue on $\pge$ and $\nge$ at KSNL. 
Projects on improvement of electronics 
and sub-noise-edge analysis\cite{pcgerandd}
are being pursued.
Dedicated dark matter experiment CDEX
with sub-keV germanium detectors are 
taking data at the new
China Jinping Underground Laboratory\cite{cjplnews}.
This facility provides the attractive features such as
a rock overburden exceeding 2400~m and horizontal 
drive-in access. 

This work is supported by
the Academia Sinica Investigator Award 2011-15,
contracts 99-2112-M-001-017-MY3
from the National Science Council, Taiwan,
and 108T502 from TUB\.{I}TAK, Turkey.


\begin{thebibliography}{99}
\bibitem{cdmpdg12}
M. Drees and G. Gerbier,
{\rm Review of Particle Physics}
Phys. Rev. {\bf D 86}, 289 (2012),
and references therein.
\bibitem{ulege}
Q. Yue et al.,  {\rm High Energy Phys. and Nucl. Phys.} 
{\bf 28}, 877 (2004); 
H.T. Wong et al.,  {\rm J. Phys. Conf. Ser.} {\bf 39}, 266 (2006).
\bibitem{ppcge}
P.N. Luke et al., {\rm IEEE Trans. Nucl. Sci.} {\bf 36}, 926 (1989);
P.A. Barbeau, J.I. Collar and O. Tench,
{\rm JCAP} {\bf 09}, 009 (2007).
\bibitem{texonocdm07}
H.T. Wong, {\rm  Mod. Phys. Lett.} {\bf A 23}, 1431 (2008);
S.T. Lin et al., {\rm Phys. Rev.} {\bf D 79}, 061101(R) (2009).
\bibitem{cogent}
C.E. Aalseth et al.,  Phys. Rev. Lett. {\bf 101}, 251301 (2008);
C.E. Aalseth et al.,  Phys. Rev. Lett. {\bf 106}, 131301 (2011);
C.E. Aalseth et al.,  Phys. Rev. Lett. {\bf 107}, 141301 (2011);
C.E. Aalseth et al.,  arXiv:1208.5737 (2012).
\bibitem{lowimptheory}
D. Hooper, Phys. Dark Univ. {\bf 1}, 1 (2012); 
C. Kelso, D. Hooper, and M.R. Buckley,
Phys. Rev. {\bf D 85}, 043515 (2012),
and references therein.
\bibitem{cdmsxenonlow}
D.S.~Akerib et al., Phys. Rev. {\bf D 82}, 122004 (2010);
Z. Ahmed et al., Phys. Rev. Lett. {\bf 106}, 131302 (2011); 
J. Angle et al., Phys. Rev. Lett. {\bf 107}, 051301 (2011);
E. Aprile et al., Phys. Rev. Lett.  {\bf 109}, 181301 (2012);
Z. Ahmed et al., arXiv:1203.1309 (2012).
\bibitem{cogentreply}
J.I.~Collar, arXiv:1010.5187 (2010);
arXiv:1103.3481 (2011);
arXiv:1106.0653 (2011);
J.I.~Collar and N.E.~Fields, arXiv:1204.3559 (2012).
\bibitem{ksnlneutrino}
H.B.~Li et al., Phys. Rev. Lett. {\bf 90}, 131802 (2003);
H.T.~Wong et al., Phys. Rev. {\bf D 75}, 012001 (2007);
M. Deniz et al., Phys. Rev. {\bf D 81}, 072001 (2010).
\bibitem{gesurface}
U. Tamm, W. Michaelis, and P. Coussieu,
Nucl. Instrum. Meth. {\bf 48}, 301 (1967);
M.G. Strauss and R.N. Larsen, 
Nucl. Instrum. Meth. {\bf 56}, 80 (1967);
E. Sakai, IEEE Trans. Nucl. Sci. {\bf 18}, 208 (1971).
\bibitem{pcgerandd}
H.T. Wong,
{\rm Int. J. Mod. Phys.} {\bf D 20}, 1463 (2011).
\bibitem{ppcmj12}
E.~Aguayo et al., 
Nucl. Instrum. Meth. {\bf A 701}, 176 (2013).
\bibitem{bpoisson}
C. Savage et al., JCAP {\bf 04}, 010 (2009).
\bibitem{geqf}
S.T. Lin et al., arXiv:0712.1645v4 (2007).
\bibitem{keycdmexpt}
R. Bernabei et al., Eur. Phys. J. {\bf C 67}, 39 (2010);
M. Felizardo et al., Phys. Rev. Lett. {\bf 108}, 201302 (2012);
G. Angloher et al., Eur. Phys. J. {\bf C 72}, 1971 (2012);
S. Archambault et al., Phys. Lett. {\bf B 711}, 153 (2012).
\bibitem{cjplnews}
K.J. Kang et al.,  {\rm J. Phys. Conf. Ser.} {\bf 203}, 012028 (2010);
Q.~Yue and H.T.~Wong, J. Phys. Conf. Ser. {\bf 375}, 042061 (2012).
\end{thebibliography}
\end{document}